\documentstyle[pra,preprint,aps,epsf]{revtex}
\begin{document}
\draft
\preprint{\vbox{\it  \null\hfill\rm DOE/ER/40561-264-INT96-00-126}\\\\}
\title{Multiple electromagnetic electron positron pair production \\ 
in relativistic heavy ion collisions}
\author{Adrian Alscher,$^{1}$ Kai Hencken,$^{1,2}$ Dirk Trautmann,
$^{1}$ and Gerhard Baur$^{1,3}$}
\address{$^{1}$Institut f\"ur Theoretische Physik, Universit\"at 
Basel, 4056 Basel, Switzerland \\
$^{2}$National Institute for Nuclear Theory, University of 
Washington, Seattle, WA 98195, USA\\ 
$^{3}$Institut f\"ur Kernphysik (Theorie), Forschungszentrum
J\"ulich, 52425 J\"ulich, Germany}
\date{\today}
\maketitle 
\tighten
\begin{abstract}
We calculate the cross sections for the production of one and more
electron-positron pairs due to the strong electromagnetic fields in
relativistic heavy ion collisions.  Using the generating functional
of fermions in an external field we derive the $N$-pair
amplitude. Neglecting the antisymmetrisation in the final state we
find that the total probability to produce $N$ pairs is a Poisson
distribution. We calculate total cross sections for the production
of one pair in lowest order and also include higher-order
corrections from the Poisson distribution up to third order.
Furthermore we calculate cross sections for the production of up to
five pairs including corrections from the Poisson distribution.
\end{abstract}
%34.90.+q Other topics in atomic and molecular collision processes 
%         and interactions
%12.20.-m Quantum electrodynamics
%11.80.-m Relativistic scattering theory
\narrowtext
\section{Introduction}

Currently new accelerators are being built at CERN (``Large Hadron
Collider'', LHC) and at Brookhaven (``Relativistic Heavy Ion
Collider'', RHIC), which allow heavy ions up to lead and uranium to
be accelerated to highly relativistic energies. The main motivation
for this is to advance into regions of baryonic densities and
temperatures in order to observe the formation of the quark gluon
plasma. But this has also renewed the interest in studying pure
electromagnetic processes in these collisions, especially the
electron-positron-pair production in peripheral collisions, where
both ions do not interact with each other.

On the one hand pair production in central collisions is a signal
one wants to use in order to study the properties of the quark gluon
plasma \cite{Vogt,Kam}. The pure electromagnetic pairs are a large
background process to these. On the other hand these processes can
also be studied already in existing accelerators (``Super Proton
Synchroton'', SPS) \cite{exp}.  Recently anti-hydrogen atoms were
produced at LEAR (``Low Energy Antiproton Ring'') in a similar
process, where antiprotons in collisions with Xenon produce
electron-positron pairs, where the positron is not produced as a
free particle, but is captured in a bound state of the antiproton
\cite{gb2}.

The theoretical treatment of this problem goes back to the beginning
of QED \cite{Landau}. But it was remarked only recently that
calculations in lowest order perturbation theory violate unitarity
\cite{bert}. In the meantime several authors have studied this
problem and found that this means that the production of multiple
pairs becomes important then. All found that the $N$-pair creation
probability can be written approximately in the form of a Poisson
distribution, which then restores unitarity
\cite{Bau,Rhodes,Best}. It was shown that generally the $N$-pair
amplitude can be reduced to the product of the vacuum amplitude and
the antisymmetrised product of $N$ reduced one-pair amplitudes. The
neglect of exchange terms in the calculation of the probability then
leads to a Poisson distribution \cite{kai3}. Corrections to this
Poisson distribution were calculated in the case of two-pair
production and were found to be small (around 1\%), its importance
decreasing also with higher Lorentz factors.

Whereas large impact parameter contribute to the single-pair
production, the $N$-pair production with increasing $N$ is dominated
mainly by small impact parameters, smaller than the Compton
wavelength of the electron.  Thus the
``equivalent-photon-approximation'' (EPA) cannot be used anymore but
exact calculations of the reduced probability are needed, since the
momentum of the virtual photon is not small compared to the electron
mass.

In Sec. II we show how the $N$-pair production amplitude can be
found using the path integral formalism. We find again the results
of \cite{kai3}. We then discuss in Sec. III our approach to the
calculation of the cross sections for one- and multiple-pair
production taken into account only peripheral collisions. Results
for this are given in Sec. IV and compared with EPA.  Differential
cross section for the one-pair production are shown too.

\section{$N$-pair production in an external field} 
In the path integral formalism the generating functional for a
fermion field $\psi$ in a external field $A_{\mu}$ can be written as
\cite{ryd}
\begin{eqnarray}
Z[\eta,\bar\eta,A]&=&N\int {\cal D}\psi{\cal D}\bar\psi\,
\exp{\biggl\{i\int d^{4}x \Bigl[
{\cal L}(x)\Bigr.\biggr.}  \nonumber \\
&& + {\biggl.\Bigl.\bar\eta(x)\psi(x)+\bar\psi(x)\eta(x)\Bigr]
\biggr\}},
\label{eqi49}
\end{eqnarray}
with the Lagrangian
\begin{equation}
{\cal L}(x)=\bar\psi(x)(i\not\!\!D -m)\psi(x),
\label{eqi50}
\end{equation}
where $D_{\mu}=\partial_{\mu}+ieA_{\mu}$ is the usual gauge
invariant derivative. We normalize $Z[\eta,\bar\eta,A]$ so that it
corresponds to the identity in the absence of the fermionic sources
and external field.  We separate the fields $\psi=\psi_{cl}+\psi'$
and $\bar\psi=\bar\psi_{cl}+\bar\psi'$ into their classical parts
and fluctuations around them. Using Feynman boundary conditions in
order to invert the equations of motion the classical solutions are
given by
\begin{eqnarray}
\psi_{cl}(x)&=&-\int d^{4}y\,S_{A}(x,y)\eta(y)\nonumber\\
\bar\psi_{cl}(x)&=&-\int d^{4}y\,\bar\eta(y)S_{A}(y,x),
\label{eqi52}
\end{eqnarray}
where $S_{A}$ is the Dirac propagator in the external field 
\begin{equation}
(i\not\!\partial-e\not\!\!A-m)S_{A}(x,y)=\delta^{4}(x-y). 
\label{eqi53}
\end{equation}
With this we can rewrite the functional in Eq. (\ref{eqi49}) in the 
form 
\begin{eqnarray}
Z[\eta,\bar\eta,A]&=&\frac{\det{[i(\not\!\!D-m)]}} 
{\det{[i(\not\!\partial-m)]}}\nonumber\\
&& \times 
\exp{\left\{-i\int d^{4}x 
d^{4}y\,\bar\eta(x)S_{A}(x,y)\eta(y)\right\}}.\nonumber\\
&&\label{eqi56}
\end{eqnarray}
The Green functions of the Dirac fields $\psi$ and $\bar\psi$ can
now be found by varying $Z$ with respect to the sources
\begin{eqnarray}
\tau(x_{1},y_{1},\ldots,x_{n},y_{n}) 
&=& \frac{1}{(i^{2})^{n}}
\frac{\delta}{\delta\eta(y_{1})}
\frac{\delta}{\delta\bar\eta(x_{1})}\ldots \nonumber\\
&&\frac{\delta}{\delta\eta(y_{n})}
\frac{\delta}{\delta\bar\eta(x_{n})}
\Biggl.Z[\eta,\bar\eta,A] \Biggr|_{\eta=\bar\eta=0}.\nonumber \\
&& \label{eq10b}
\end{eqnarray}
We use the asymptotic $out$-system. The initial and final states for 
$N$-pair creation are then
\begin{eqnarray}
\left|i,\,out\right\rangle &=&\left|0\right\rangle \nonumber\\
\left|f,\,out\right\rangle &=& d^{+}_{out\,s_{1}}(p_{1})
b^{+}_{out\,s_{1}'}(p_{1}') 
\ldots \nonumber\\
&& d^{+}_{out\,s_{N}}(p_{N})b^{+}_{out\,s_{N}'}(p_{N}')
\left|0\right\rangle ,
\label{eq15b}
\end{eqnarray}
where the one-particle creation operators $b^+$ and $d^+$ obey the
usual anticommutation relations for free particles.  In order to get
the amplitude for the creation we make use of the
Leh\-mann-Syman\-zik-Zimmer\-mann reduction \cite{LSZ}. The first
term in Eq. (\ref{eqi56}) can be identified with the vacuum
amplitude \cite{kai3}. Thus we get the amplitude in the form
\begin{eqnarray}
\left\langle f|S|i\right\rangle &&=\left\langle 0|S|0\right\rangle 
\int d^{4}x_{1}\ldots d^{4}x_{N}d^{4}y_{1}\ldots d^{4}y_{N} 
\nonumber\\
\biggl\{\biggr.\prod_{j=1}^{N}&&\bar u^{(s_{j}')}
(p_{j}')e^{ip_{j}'x_{j}}(i\not\!{\roarrow{\partial}}_{x_{j}}-m)
 \tau(x_{1},y_{1},\ldots,x_{N},y_{N})\nonumber\\
\biggl.\prod_{k=1}^{N}&&(-i\not\!{\loarrow{\partial}}_{y_{k}}-m)
e^{ip_{k}y_{k}} v^{(s_{k})}\biggr\}.
\label{eq36b}
\end{eqnarray}
The variations in Eq. (\ref{eq10b}) lead only to connected
graphs. We extract the individual fermion lines with the help of the
Wick theorem \cite{ItZ} in order to get
\begin{equation}
\left\langle f|S|i\right\rangle  = \left\langle 0|S|0\right\rangle 
\sum_{\sigma\in S_{N}} \epsilon_{\sigma}
S^{R}(u_{1},v_{\sigma(1)})\ldots S^{R}(u_{N},v_{\sigma(N)}),
\label{eq43b}
\end{equation}
where we have summarized momentum and spin quantum numbers of
electrons as $u_{i}=(p_{i}',s_{i}')$ and of positrons as
$v_{j}=(p_{j},s_{j})$ and summed over all permutations of $S_{N}$.
The reduced one-pair amplitude $S^{R}(u_{i},v_{j})$ describes a
single fermion line interacting with the external field to arbitrary
order. They are given by expanding the Dirac Propagator $S_{A}$ and
doing the integrals in Eq. (\ref{eq36b}), see also \cite{kai3},
\begin{eqnarray}
S^{R}(u_{i},v_{j}) &=& 
i\sum_{n=2}^{\infty}\Biggl[
 \int d^{4}z_{1}\ldots d^{4}z_{n}\,
\bar u^{(s_{i}')}(p_{i}')e^{ip_{i}'z_{n}}\Biggr.\nonumber\\
& & \times  \biggl\{
 \left[e\not\!\!A(z_{n})\right]S_{F}(z_{n}-z_{n-1})
\ldots \left[e\not\!\!A(z_{1})\right]\biggr\}  \nonumber\\
&&\Biggl. \times e^{ip_{j}z_{1}}v^{(s_{j})}(p_{j})\Biggr].
\label{eq42b}
\end{eqnarray}
The $N$-pair amplitude in Eq. (\ref{eq43b}) is therefore the product
of the vacuum amplitude and an antisymmetrised product of the
reduced amplitudes.  Due to the correlation between the produced
electrons and positrons it is justified to neglect the Pauli
principle in the final state, i.e., terms in the probability for $N$
pairs coming from two different permutations are not taken into
account. This leads finally to the probability to produce $N$ pairs
\begin{equation}
P(N)=\frac{\left[P^{R}\right]^{N}}{N!} e^{-P^{R}},
\label{eq48b}
\end{equation}
where the vacuum amplitude shows up as the exponential function.  As
shown in \cite{kai3} the neglect of the exchange terms does not give
rise to large deviations for total two-pair creation
probabilities. Therefore the use of the Poisson distribution seems
to be justified. With the knowledge of reduced one-pair
probabilities, Eq. (\ref{eq48b}) allows the calculation of $N$-pair
probabilities and $N$-pair cross sections.

\section{Calculation of cross sections}
The first order that contributes to the reduced one-pair amplitude
is the second order. For the creation of an electron with momentum
$p_{-}$ and spin $s_{-}$ and a positron with momentum $p_{+}$ and
spin $s_{+}$ we get the amplitude \cite{kai2}
\begin{eqnarray}
S_{fi}^{R}&=&\bar u^{(s_{-})}(p_{-}) 
ie^{2}\biggl\{\int \frac{d^{4}p}{(2\pi)^{4}}\not\!\!\tilde{A}(p_{-}-p)
\frac{\not\!p+m}{p^{2}-m^{2}}\biggr. \nonumber\\
&& \times \biggl.\not\!\!\tilde{A}(p_{+}+p)\biggr\}v^{(s_{+})}(p_{+}).
\label{eq28c}
\end{eqnarray}
We describe the electromagnetic potential of the two ions as a  
superposition of the four-potentials 
\begin{eqnarray}
\tilde{A}_{\mu}(q)&=& -2\pi Ze\frac{F(q)}{q^{2}}\biggl\{
u_{\mu}^{(1)}e^{iq\frac{b}{2}} 
\delta\left(qu^{(1)}\right)\biggr. \nonumber\\
&& + \biggl.u_{\mu}^{(2)}e^{-iq\frac{b}{2}} \delta\left(qu^{(2)}
\right)\biggr\},
\label{eq30a}
\end{eqnarray}
using the straight-line approximation for the trajectories of the
ions.  $b$ is the impact parameter, $u^{(1)}=\gamma
(1,0,0,\beta)=\gamma w^{(1)}$ and $u^{(2)}=\gamma
(1,0,0,-\beta)=\gamma w^{(2)}$ are the four-velocities of the ions
in the center-of-velocity frame.  Throughout the calculations we are
making use of the monopole form factor
\begin{equation}
F(q)=\frac{\Lambda^{2}}{\Lambda^{2}-q^{2}},
\label{eqdip}
\end{equation}
with $\Lambda=83\,MeV$ for $Au$ ions, in order to describe the
extended charge distribution of the ion, as this form factor can be
treated analytically \cite{kai1}.  One-pair cross sections are not
sensitive to the detailed choice of the form factor. The cross
sections for the multiple-pair productions are more sensitive to
this as they are produced mainly at smaller impact parameters.

We get the total reduced probability by summing the absolute value
squared of Eq. (\ref{eq28c}) over all electron and positron spins
and integrating over all momenta
\begin{equation}
P^{R}(b)=\int \frac{d^{3}p_{-}d^{3}p_{+}}
{(2\pi)^{6}}\frac{m^{2}}{E_{-}E_{+}}\sum_{s_{-},s_{+}}
|S_{fi}^{R}|^{2}.
\label{eq40c}
\end{equation}
By rewriting the spin summation as a Dirac trace over
$\gamma$-matrices we get
\widetext
\begin{eqnarray}
\sum_{s_{-},s_{+}}|S_{fi}^{R}|^{2}&=&(Z\alpha)^{4}
\frac{4}{\beta^{2}}
 tr\Biggl[\int d^{2}q_{1\bot}\,
\frac{1}{q_{1}^{2}[q_{1}-p_{+}-p_{-}]^{2}} e^{iq_{1}b} \Biggr.
\nonumber\\ && \times
\biggl\{
 \frac{\not\!\!w^{(1)}(\not\!p_{-}-\not\!q_{1} +m)\not\!\!w^{(2)}}
{[(q_{1}-p_{-})^{2}-m^{2}]}
+ \frac{\not\!\!w^{(2)}(\not\!q_{1}-\not\!p_{+} +m)\not\!\!w^{(1)}}
{[(q_{1}-p_{+})^{2}-m^{2}]}
\biggr\}\frac{\not\!p_{+}-m}{2m}\nonumber\\
&& \times \int d^{2}q'_{1\bot}\, 
\frac{1}{q_{1}'^{2}[q_{1}'-p_{+}-p_{-}]^{2}}
e^{-iq_{1}'b}
\nonumber\\ && \times
\Biggl.\biggl\{
 \frac{\not\!\!w^{(2)}(\not\!p_{-}-\not\!q_{1}' +m)\not\!\!w^{(1)}}
{[(q_{1}'-p_{-})^{2}-m^{2}]}
+ \frac{\not\!\!w^{(1)}(\not\!q_{1}'-\not\!p_{+} +m)\not\!\!w^{(2)}}
{[(q_{1}'-p_{+})^{2}-m^{2}]}\biggr\}\frac{\not\!p_{-}+m}{2m}\Biggr],
\label{eq45c}
\end{eqnarray}
\narrowtext
\noindent
where $tr$ is the usual trace over the Dirac indices.  The $\delta$
functions in the amplitude in Eq. (\ref{eq28c}) determine the
longitudinal components of the integration variables.  We still have
to integrate over the components transverse to the velocity of the
ions. We have changed these variables to the momentum of one of the
photons $q_{1}=p_{-}-p$ and to the difference $q=q_{1}'-q_{1}$.
This allows us to rewrite the $b$-dependent probability in
Eq. (\ref{eq40c}) in the form
\begin{equation}
P^{R}(b)=\int d^{2}q\tilde{P}^{R}(q)
e^{i\vec{q}\,\vec{b}},
\label{eq57c}
\end{equation}
\noindent
where the Fourier transformed probability is
\widetext
\begin{eqnarray}
\tilde{P}^{R}(q)&=&\int\frac{d^{3}p_{-}d^{3}p_{+}}
{(2\pi)^{6}E_{-}E_{+}}\frac{(Z\alpha)^{4}}{\beta^{2}} 
tr\Biggl[\int d^{2}q d^{2}q_{1}\,\frac{1}{N_{0}N_{1}N_{3}N_{4}} 
\Biggr.
\nonumber\\
&& \times
\biggl\{
 \frac{\not\!\!w^{(1)}(\not\!p_{-}-\not\!q_{1} +m)
\not\!\!w^{(2)}}{N_{2D}}
+ \frac{\not\!\!w^{(2)}(\not\!q_{1}-\not\!p_{+} +m)
\not\!\!w^{(1)}}{N_{2X}}
\biggr\}(\not\!p_{+}-m)\nonumber\\
&&\times \Biggl. \biggl\{
 \frac{\not\!\!w^{(2)}(\not\!p_{-}-\not\!q_{1}' +m)
\not\!\!w^{(1)}}{N_{5D}}
+ \frac{\not\!\!w^{(1)}(\not\!q_{1}'-\not\!p_{+} +m)
\not\!\!w^{(2)}}{N_{5X}}
\biggr\}(\not\!p_{-}+m)\Biggr],\label{eq56c}
\end{eqnarray}
\narrowtext
\noindent
with the propagators
\begin{equation}
N_{0}=\vec{q}_{1}^{\,2}+m_{0}^{2},\quad 
N_{i}=(\vec{q}_{1}+\vec{k}_{i})^{2}+m_{i}^{2}.
\label{eq11d}
\end{equation}
The momenta $\vec{k}_{i}$ and parameters $m_{i}^{2}$ of the
propagators in Eq. (\ref{eq56c}) are defined as (see also
\cite{kai2})
\begin{equation}
\begin{array}{l c l}
& \quad & m_{0}^{2}= -q_{1l}^{2}, \\
\vec{k}_{1}=-\vec{p}_{+}-\vec{p}_{-}&&
m_{1}^{2}=-(q_{1l}-p_{+l}-p_{-l})^{2},\\
\vec{k}_{2D}=-\vec{p}_{-} &&
m_{2D}^{2}=m^{2}-(q_{1l}-p_{-l})^{2},\\
\vec{k}_{2X}=-\vec{p}_{+} &&
m_{2X}^{2}=m^{2}-(q_{1l}-p_{+l})^{2},\\
\vec{k}_{3}=\vec{q} &&
m_{3}^{2}=m_{0}^{2},\\
\vec{k}_{4}=\vec{q}-\vec{p}_{+}-\vec{p}_{-} &&
m_{4}^{2}=m_{1}^{2},\\
\vec{k}_{5D}=\vec{q}-\vec{p}_{-} &&
m_{5D}^{2}=m_{2D}^{2},\\
\vec{k}_{5X}=\vec{q}-\vec{p}_{+} &&
m_{5X}^{2}=m_{2A}^{2}.
\end{array}
\label{eq46c}
\end{equation}
The trace over the Dirac $\gamma$-matrices is done with the
algebra-program FORM \cite{FORM}.  The advantage of calculating the
Fourier transform of the total probabilities is due to the absence
of oscillating terms in Eq. (\ref{eq57c}), which would otherwise
prevent convergence in the Monte Carlo integrations.

From this we can get the total cross sections $\sigma$ by
integrating the probabilities (Eq. (\ref{eq48b})) over the impact
parameter
\begin{equation}
\sigma(N)=2\pi\int_{2R}^{\infty}db\,b P(N,b),
\label{eq1d}
\end{equation}
starting with $2R$ (for symmetric collisions) in order to take only
peripheral collisions into account.

We can use $P(N=1,b)$ directly as given in Eq. (\ref{eq57c}). But
for large impact parameter the integrand oscillates very fast
allowing not a very accurate calculation of $P$. Therefore we choose
to do the calculation for the cross section for the one-pair
production by expanding the Poisson distribution of $\sigma(N=1)$ up
to third order
\begin{equation}
\sigma(N=1)=\sigma^{(1)} -\frac{2!}{1!}\sigma^{(2)} 
+\frac{3!}{2!}\sigma^{(3)}-\cdots . 
\label{eq4d}
\end{equation}
Integrating from $b=0$ on the $\sigma^{(n)}$ can be written as
\begin{mathletters}
\label{eq8d}
\begin{eqnarray}
\sigma^{(1)}_{b\geq 0}&=&\int d^{2}b d^{2}q\, 
e^{i\vec{q}\,\vec{b}}\tilde{P}^{R}(q)\nonumber\\
&=& (2\pi)^{2}\int d^{2}q \, \delta^{2}\left(\vec{q}\right)
\tilde{P}^{R}(q)\nonumber\\
&=&(2\pi)^{2}\tilde{P}^{R}(0), \label{eq8da}
\end{eqnarray}
\begin{equation}
\sigma^{(2)}_{b\geq 0}=\frac{(2\pi)^{3}}{2!}\int 
dq\,q\left[\tilde{P}^{R}(q)\right]^{2}, \label{eq8db}
\end{equation}
\begin{equation}
\sigma^{(3)}_{b\geq 0}=\frac{(2\pi)^{2}}{3!}
\int d^{2}q_{1}d^{2}q_{2}\, \tilde{P}^{R}(q_{1})\tilde{P}^{R}(q_{2})
\tilde{P}^{R}(|\vec{q}_{1}+\vec{q}_{2}|).\label{eq8dc}
\end{equation}
\end{mathletters}
\noindent
Therefore we can calculate the cross section by using the Fourier
transformed $\tilde{P}^{R}(q)$ alone. This leads to more accurate
results for the cross section as the cross section for the one-pair
production is dominated by large impact parameter, where the impact
parameter dependent probability is not very accurate due to the
large number of oscillations in the Fourier transform.

For the calculation of $\sigma^{(1)}$ we have to set $\vec{q}$ to
zero in Eq. (\ref{eq56c}), i.e. $\vec{q}_{1}\,'=\vec{q}_{1}$. This
leads to a number of simplifications
\begin{equation}
N_{3}=N_{0},\, N_{4}=N_{1},\,N_{5D}=N_{2D},\,N_{5X}=N_{2X}.
\label{eq10d}
\end{equation}

The integrations over $\vec{q}_{1}$ can then be done analytically as
they are standard two-dimensional Feynman integrals. The
$\vec{q}_{1}$'s in the numerator only appear in connection with the
scalar products $\vec{p}_{+}\vec{q}_{1},\,\vec{p}_{-}\vec{q}_{1}$
and $\vec{q}_{1}\vec{q}_{1}$ and the $\vec{k}_{i}$'s in the
propagators are also either $-\vec{p}_{-},\,-\vec{p}_{+}$ or their
sum.  Therefore by using the relations
\begin{eqnarray}
\vec{q}_{1}^{\,2}&=&N_{0}^{2}-m_{0}^{2}\nonumber\\
2\vec{q}_{1}\vec{k}_{i}&=&N_{i}-N_{0}- r_{i}\label{eq12d},
\end{eqnarray}
we can express the scalar products in the numerator as propagators
and scalars in order to eliminate all $\vec{q}_{1}$'s from the
numerator. In the end we are left with scalar Feynman integrals with
different numbers of propagator terms in the denominator. The higher
integrals with quadratic propagator terms cannot be reduced to
simpler integrals with the method described in \cite{kai2,NV}. But
we solve them by differentiating the standard integrals with respect
to the parameters $m_{0}^{2}$ and $m_{i}^{2}$.

The integration over the momenta of electron and positron in the
final state are done with the Monte Carlo integration routine VEGAS
\cite{Vegas}.  We write the momenta in polar coordinates in the
plane transverse to the velocity of the ions. The integration over
one of the angles is trivial and there remains a five dimensional
integral over momenta and the relative angle between the transverse
electron and positron momenta. The integration boundaries were
incremented until the Monte Carlo integration converged with
accuracies better than 0.5\%. We did not take any form factors into
account for $\tilde{P}^{R}(0)$'s as the results of \cite{kai2} show
that the form factor is only important for small $b$'s, i.e., only
for large $q$'s. At small $q$'s differences between calculations
with and without form factors are not observable.

For the second order correction $\sigma^{(2)}$ (Eq. (\ref{eq8db}))
we calculate the values of $\tilde{P}^{R}(q)$ for $q\not=0$ with the
method described in \cite{kai2,kai1}. We are adding to this the
value at $q=0$, and approximate the $\tilde{P}^{R}(q)$ by cubic
splines.  We use a logarithmic scale for $q$ for the interpolation
in order to get a better approximation for the steep increase at
small $q$'s.

The calculation of the third order correction $\sigma^{(3)}$ 
(Eq. (\ref{eq8dc})) follows that of $\sigma^{(2)}$. 

Finally we have to correct for the fact, that we are integrating
from $b=0$ instead of $2R$. First we calculate the impact parameter
dependent probability (Eq. (\ref{eq57c})) by integrating over the
angular part and get
\begin{equation}
P^{R}(b)=2\pi\int dq\,q\tilde{P}^{R}(q)J_{0}(qb),
\label{eq16d}
\end{equation}
with the Bessel function $J_{0}$. Then we subtract from the
$\sigma^{(n)}$ the contribution from $0<b<2R$ by integrating
\begin{equation}
\frac{d\sigma^{(n)}}{db}=2\pi b\frac{[P^{R}(b)]^{n}}{n!},
\label{eqneu1}
\end{equation}
from $b=0$ to $2R$. As $P^{R}(b)$ only decreases slowly with $b$,
large impact parameters are still important for the cross
section. The Bessel function in Eq. (\ref{eq16d}) oscillates fast
for large $b$, which would lead to a poor accuracy for
$\sigma^{(1)}$ if one would integrate over the impact parameter
directly.

For increasing $N$ pair production occurs within increasingly
smaller $b$'s as the external field has to supply the minimal energy
of $2Nmc^{2}$ to create $N$ pairs, which further restricts the
system. In this case the influence of the large $b$'s decreases and
we can calculate the total multiple-pair cross sections exactly by
using the whole Poisson distribution and integrating directly over
the impact parameters
\begin{equation}
\sigma(N)=2\pi\int_{2R}^{\infty} db\,b\frac{[P^{R}(b)]^{N}}{N!}
e^{-P^{R}(b)}.
\label{eqneu2}
\end{equation}

\section{Results and Conclusions}
\begin{figure}
\centerline{\epsfxsize=8.6cm \epsffile{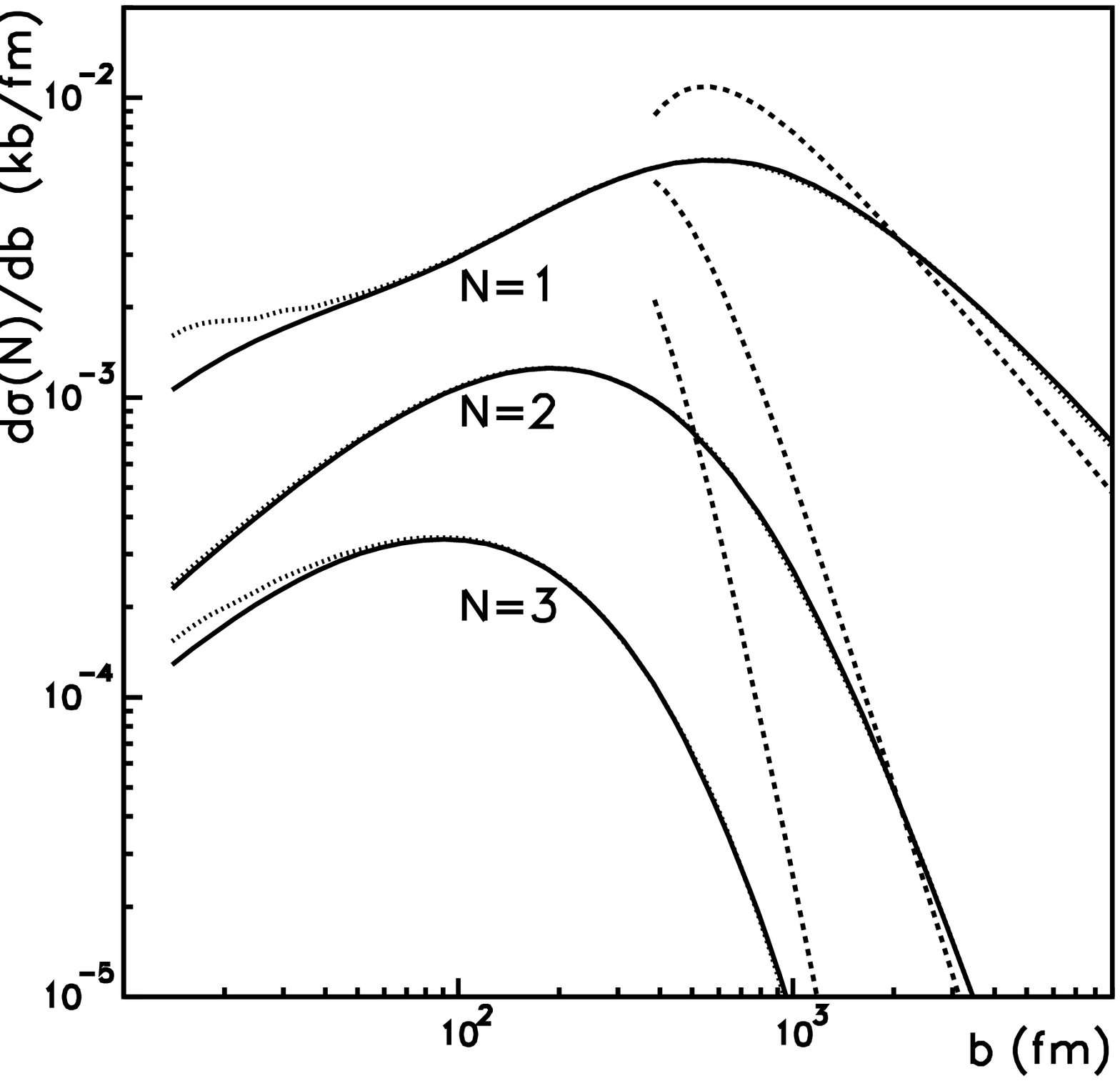}}
\caption{The $b$ dependent differential cross sections
$\frac{d\sigma(N)}{db}(b)$ for the $N$-pair production in a
$^{197}Au^{79}-^{197}Au^{79}$ collision with $\gamma=100$ for
$N=1,2,3$.  The solid lines show the results for the calculation
with a monopole form factor, the dotted lines for a point charge and
the dashed lines are the EPA results.}
\label{fig1}
\end{figure}

Using the method described in the last section we have calculated
the impact parameter dependence of the one- and multiple-pair cross
section up to three pairs. In Fig.  \ref{fig1} we show the results
of the calculation with and without a monopole form factor for a
center-of-mass Lorentz factor $\gamma=100$ and for a
$^{197}Au^{79}-^{197}Au^{79}$ collision.  Also shown are the results
of the equivalent photon approximation (EPA) using for the reduced
one-pair probability in the Poisson distribution \cite{bert}
\begin{equation}
P^{EPA}(b)=\frac{14}{9\pi^{2}}(Z\alpha)^{4}
\left(\frac{\lambdabar_{c}}{b}\right)^{2}
\log^{2}{\left(\frac{\lambdabar_{c}\gamma_{tar}\delta}{2b}\right)}
\label{eq25d},
\end{equation}
with $\delta=0.681$. $\lambdabar_{c}$ is the Compton wavelength of
the electron and $\gamma_{tar}=2\gamma^{2}-1$ is the Lorentz factor
in the target system. The EPA approximation of Eq. (\ref{eq25d}) is
only valid for impact parameters
$\lambdabar_{c}\gamma_{tar}\delta\geq 2b \geq 2\lambdabar_{c}$.  We
find in agreement with earlier results \cite{kai2,kai1} that the EPA
gives results which are much to large at small $b$. As already
discussed before, the one-pair production cross section decreases
only slowly with larger impact parameter in contrast to the
multiple-pair cross sections.

\begin{figure}
\centerline{\epsfxsize=8.6cm \epsffile{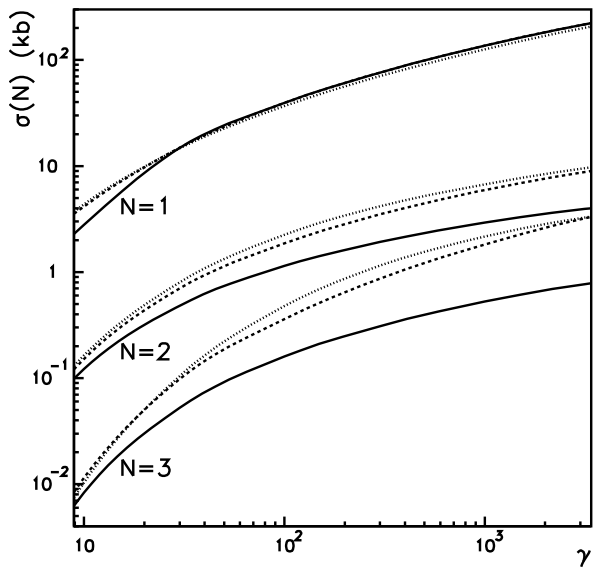}}
\caption{The $\gamma$ dependent total cross sections
$\sigma(N)(\gamma)$ for the $N$-pair production in a
$^{208}Pb^{82}-^{208}Pb^{82}$ collisions, for $N=1,2,3$.  Exact
calculations are shown as solid lines, dashed lines are the results
for the corresponding first orders $\sigma^{(N)}(\gamma)$ and the
dotted lines are the EPA results.}
\label{fig2}
\end{figure}

In Fig. \ref{fig2} we show the dependence of the total cross section
for the one-, two- and three-pair production for a $^{208}Pb^{82} -
^{208}Pb^{82}$ collision on the Lorentz factor $\gamma$. We plot the
results for the calculation with a monopole form factor using the
full Poisson distribution and also using the lowest order alone. One
sees that the Poisson distribution reduces the multiple-pair cross
sections considerably whereas it seems to have only a small effect
on the one-pair production. Also shown are the results of the EPA
approximation by integrating only over impact parameters larger than
the Compton wavelength. They are in fair agreement with the exact
results, even though they tend to be either to large or to small
systematically.

Our results for the total cross sections are summarized for the
different colliders in Tables \ref{tab1} and \ref{tab2}. The
differences between the calculations with and without a monopole
form factor show the dependence on the detailed form of the charge
distribution in the ion. More realistic form factors are needed in
order to describe pair production with a larger $N$ more
accurately. Moreover our results are in agreement with the results
of \cite{gue}, which were calculated only at lower $\gamma$ and with
a Monte Carlo integration alone.

Recently anti-hydrogen atoms were produced in a LEAR experiment at
CERN by colliding antiprotons with Xenon \cite{gb2} with
$\gamma_{tar}=1.8$.  With our approach we get a total cross section
for the creation of a free electron-positron pair in this case of
$\sigma^{(1)}=Z^{2}\sigma_{0}$, with $\sigma_{0}=14 \, nb$. This is
more than three orders of magnitude larger than the bound-free
transition, which leads to anti-hydrogen formation.

We have applied our calculations only to electron-positron-pair
production. Nevertheless the pair production of heavy leptons $\mu$
and $\tau$ gets more important with increasing energies of the
ions. With our method we can calculate cross sections for this
processes, too, considering only peripheral collisions. Moreover
only single-pair cross sections are important here due to the larger
masses.

\begin{figure}
\centerline{\epsfxsize=8.6cm \epsffile{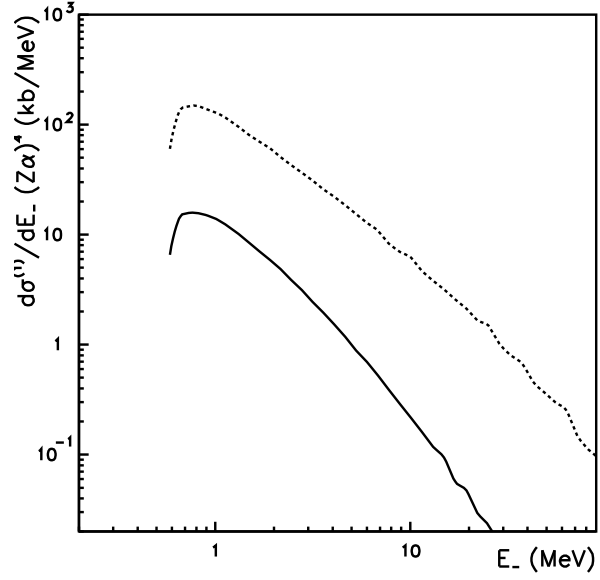}}
\caption{Differential cross sections $\frac{1}{(Z\alpha)^{4}}
\frac{d\sigma^{(1)}}{dE_{-}}(E_{-})$ for the first order in
Eq. (\protect\ref{eq4d}) as a function of the energy of the electron
or positron. The solid line shows the result for $\gamma=8.916$ and
the dashed line for $\gamma=100$.}
\label{fig3}
\end{figure}

\begin{figure}
\centerline{\epsfxsize=8.6cm \epsffile{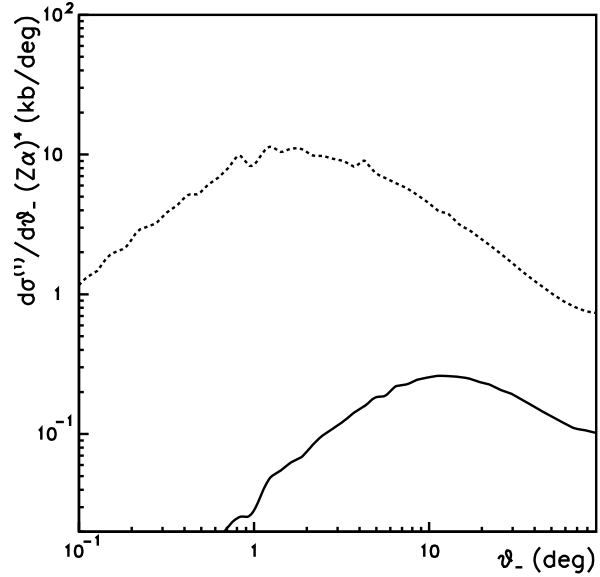}}
\caption{Differential cross sections
$\frac{1}{(Z\alpha)^{4}}
\frac{d\sigma^{(1)}}{d\vartheta_{-}}(\vartheta_{-})$
as a function of the electron momenta with the beam axis.  Same
notations as in Fig. \protect\ref{fig3}.}
\label{fig4}
\end{figure}

Finally we also give differential cross sections for the energy and
the angle with the beam axis of either electron or positron. We get
these by integrating the first order of the one-pair cross sections
in Eq. (\ref{eq8d}) with the help of the Monte Carlo integration
routine sorting the results into bins. The results are shown in
Figs. \ref{fig3} and \ref{fig4}. Most electrons are produced rather
close to the beam axis and also with relatively low energy of the
order of a few $MeV$.

\acknowledgments
This work was supported in part by the Swiss National Science
Foundation (SNF) and the ``Freiwillige Akademische Gesellschaft''
(FAG) of the University of Basel. One of us (K.H.) would like to
thank them for their financial support.

\mediumtext
\begin{table}
\begin{center}
\begin{tabular}{c c c c c c}
& \multicolumn{5}{c}{with monopole form factor (kb)}\\
$\gamma$ & $\sigma(N=1)$ & $\sigma (N=2)$  & $\sigma(N=3)$
& $\sigma(N=4)$ & $\sigma(N=5)$ \\
\hline
8.916\, (SPS) & 
2.29 & 9.89\,$10^{-2}$ & 6.26\,$10^{-3}$ & 4.52\,$10^{-4}$ 
& 3.25\,$10^{-5}$\\
100\, (RHIC)  & 
31.8 & 0.899 & 0.1122 & 1.827\,$10^{-2}$ & 3.11\,$10^{-3}$ \\
3400\, (LHC) & 
223 & 4.00 & 0.785 & 0.219 & 6.94\,$10^{-2}$ \\  
\end{tabular}
\end{center}
\caption{Results of the calculation for the production of
$N=1,\ldots,5$ pairs with a monopole form factor. For SPS and LHC we
show total cross sections for $^{208}Pb^{82}-^{208}Pb^{82}$
collisions and for RHIC total cross sections for
$^{197}Au^{79}-^{197}Au^{79}$ collisions.}
\label{tab1}
\end{table}

\begin{table}
\begin{center}
\begin{tabular}{c c c c c c}
& \multicolumn{5}{c}{without monopole form factor (kb)} \\
$\gamma$ & $\sigma(N=1)$ & $\sigma (N=2)$  & $\sigma(N=3)$
& $\sigma(N=4)$ & $\sigma(N=5)$ \\
\hline
8.916\, (SPS) & 
2.28 & 0.1019 & 6.66\,$10^{-3}$ & 5.08\,$10^{-4}$ & 3.95\,$10^{-5}$\\
100\, (RHIC)  & 
31.8 & 0.897 & 0.1136 & 1.881\,$10^{-2}$ & 3.37\,$10^{-3}$ \\
3400\, (LHC) & 
223 & 3.95 & 0.776 & 0.219 & 7.09\,$10^{-2}$ \\  
\end{tabular}
\end{center}
\caption{Results for the calculations for the production of
$N=1,\ldots,5$ pairs for a point charge.  The ions used for the
different $\gamma$'s are the same as in Tab. \protect\ref{tab1}.}
\label{tab2}
\end{table} 

\begin{references}
\bibitem{Vogt} R. Vogt {\em et. al. \/},
 Nucl. Phys. A {\bf 583}, 693c (1995).
\bibitem{Kam} B. K\"ampfer {\em et. al. \/}, 
Phys. Rev. C {\bf 52}, 2704 (1995).
\bibitem{exp} T. Alber {\em et. al.\/}, 
Phys. Rev. Lett. {\bf 75}, 3814 (1995). 
\bibitem{gb2} G. Baur {\em et. al.}, 
Phys. Lett. B {\bf 368}, 251 (1996).
\bibitem{Landau} L.D. Landau and E.M. Lifshitz, 
Phys. Z. Sowjet. {\bf 6}, 244 (1934). 
\bibitem{bert} C.A. Bertulani and G. Baur, 
Phys. Rep. {\bf 164}, 300 (1987).
\bibitem{Bau} G. Baur, 
Phys. Rev. A {\bf 42}, 5736 (1990).
\bibitem{Rhodes} M. J. Rhoades-Brown and J. Weneser, 
Phys. Rev. A {\bf 44}, 330 (1991).
\bibitem{Best} C. Best, W. Greiner, and G. Soff, 
Phys. Rev. A {\bf 46}, 261 (1992).
\bibitem{kai3} K. Hencken, D. Trautmann, and G.~Baur, 
Phys. Rev. A {\bf 51}, 998 (1995).
\bibitem{ryd} L.H. Ryder, {\em Quantum Field Theory}, 
(Cambridge University Press, 1985).
\bibitem{LSZ} H. Lehmann, K. Symanzik, and W. Zimmermann, 
Nuovo Cimento {\bf 1}, 205 (1955). 
\bibitem{ItZ} C. Itzykson and J.-B. Zuber, 
{\em Quantum Field Theory\/}, (McGraw-Hill, 1980).
\bibitem{kai2} K. Hencken, D. Trautmann, and G.~Baur, 
Phys. Rev. A {\bf 51}, 1874 (1995).
\bibitem{kai1} K. Hencken, D. Trautmann, and G.~Baur, 
Phys. Rev. A {\bf 49}, 1584 (1994).
\bibitem{FORM} FORM is an algebraic calculation program by 
J.A.M. Vermaseren; the free version 1.0 can be found, e.g. at 
FTP.NIKHEF.NL
\bibitem{NV} W.L. van Neerven and J.A.M. Vermaseren, 
Phys. Lett. B {\bf 137}, 241 (1984). 
\bibitem{Vegas} G.P. Lepage, 
J. Comp. Phys. {\bf 27}, 192 (1978).
\bibitem{gue} M.C. G\"u\c{c}l\"u {\em et. al.}, 
Phys. Rev. A {\bf 51}, 1836 (1995).
\end{references}
\end{document}